\documentclass{PoS}
\PoS{PoS(LAT2005)060}
\usepackage{graphicx}
\usepackage{amsmath,amssymb}

\def\sla#1{\rlap{\kern .13em /}#1}

\def\simge{
    \mathrel{\rlap{\raise 0.511ex
        \hbox{$>$}}{\lower 0.511ex \hbox{$\sim$}}}}
\def\simle{
    \mathrel{\rlap{\raise 0.511ex
        \hbox{$<$}}{\lower 0.511ex \hbox{$\sim$}}}}

\title{Finite Size Effect in Excited Baryon Spectroscopy\thanks{
All calculations have been carried out 
on a Hitachi SR8000 parallel computer 
at KEK (High Energy Accelerator Research Organization). 
This work is supported by the Supercomputer Projects No.102 (FY2003) 
and No.110 (FY2004) of High Energy Accelerator Research Organization (KEK). 
}}
\ShortTitle{Finite Size Effect in Excited Baryon Spectroscopy}

\author{\speaker{Kiyoshi Sasaki}\\
\noindent
Center for Computational Science, University of Tsukuba,\\
Tsukuba, Ibaraki 305-8577, Japan\\
E-mail: \email{ksasaki@ccs.tsukuba.ac.jp}}

\author{Shoichi Sasaki\\
Department of Physics, University of Tokyo,\\
Tokyo 113-0033, Japan\\
and\\
RIKEN BNL Research Center, Brookhaven National Laboratory,\\
Upton, NY 11973-5000, USA\\
E-mail: \email{ssasaki@phys.s.u-tokyo.ac.jp}}

\abstract{
We investigate the finite size effect on masses of excited baryons
in quenched lattice QCD simulation. 
For this purpose, we perform numerical simulations at
three different lattice sizes, $La\simeq 1.6$, 2.2 and 3.2 fm. 
The gauge configurations are generated 
with the single plaquette gauge action at $\beta=6/g^2=6.2$, 
and the quark propagator are computed with the Wilson fermion action. 
To access to two parity states of the nucleon and four different spin-parity states of the $\Delta$ baryon, the appropriate spin/parity projection are carried out.
We find that the spatial lattice size is required to be as large as 
3 fm to remove the finite size effect on excited baryons even 
in the heavy quark region ($M_\pi/M_\rho\simeq 0.82,~ 0.87$),
where that on the nucleon is negligible.
On our largest lattice ($La\simeq 3.2$ fm), all mass spectra of the 
$J^P=1/2^\pm$ nucleons and the $J^P=1/2^\pm, 
3/2^\pm$ $\Delta$ baryons are roughly consistent with 
experimental values after naive chiral extrapolation.}

\FullConference{XXIIIrd International Symposium on Lattice Field Theory\\
		 25-30 July 2005\\
		 Trinity College, Dublin, Ireland}

\begin{document}


Recently, many lattice studies on excited baryon spectroscopy 
have been made within the quenched approximation, 
and gave results in gross agreement with experiments
\cite{
Sasaki:2001nf,Gockeler:2001db,Melnitchouk:2002eg,
Zanotti:2003fx,Nemoto:2003ft,Brommel:2003jm,Mathur:2003zf}. 
To achieve the higher precision study, however, 
it is necessary to control several systematic errors, 
which may arise from the quenched approximation, 
chiral extrapolation, nonzero lattice spacing and finite size effect. 
In this study, we especially focus on the finite size effect, 
which has been not much paid attention so far. 

From the phenomenological point of view, 
the finite size effect 
may be interpreted 
by the ``wave function" squeezed in the finite volume 
\cite{Fukugita:1992jj}. 
Accordingly, to remove this effect, 
the wave function should be completely fitted in a finite lattice box. 
As reported in Ref. \cite{Aoki:1993gi}, 
the spatial size of lattice is required to be $La>2.5$ fm 
even for the ground-state nucleon in quenched lattice QCD. 
It is naively expected that 
the much larger size is required for the excited state 
than the ground state. However, most of the past studies are 
performed with spatial lattice sizes $La\simeq 1.6-2.2$ fm
due to large computational costs~\footnote{
In Ref.\cite{Mathur:2003zf}, a large lattice size ($La\sim 3.2$ fm)
for the simulation is achieved with a relatively coarse lattice spacing.
We however recall that masses of excited baryons are beyond 
their cutoff scale ($a^{-1}\sim 1$ GeV).
}. 

To estimate the finite size effect, 
we perform simulations with three different lattice sizes, 
$La\simeq 1.6$, 2.2 and 3.2 fm. 
We examine this effect on masses of the ground and excited baryons, 
the $J^P=1/2^\pm$ nucleons and the $J^P=1/2^\pm, 3/2^\pm$ $\Delta$ baryons. 
In general, the local baryon operator couples to 
both the positive- and negative-parity state. 
The appropriate parity projection is desired since the energy level of the 
negative-parity baryon is higher than that of the ground state.
The spin projection is required for a precise treatment of the $\Delta$ baryon spectra. 
We briefly explain the projection technique that we utilized in this study as below.



The local baryon operator 
${\cal O}_B(x)=\varepsilon_{abc}(q_a^T(x)C\Gamma q_b(x))\Gamma' q_c(x)$ 
couples to both the positive- and negative-parity state 
regardless of the intrinsic parity of the operator~\cite{Sasaki:2001nf}. 
Thus, the two-point function $G_B(t)= \sum_{\vec{x}}
\langle 0|T\{{\cal O}_B(\vec{x},t)\bar{\cal O}_B(\vec{0},0)\}|0\rangle$ 
has the following asymptotic form \cite{Sasaki:2001nf,Fucito:1982ip},
\begin{equation}
  G_B(t)=A_+\frac{1+\mathrm{sgn}(t)\gamma_4}{2}e^{-M_+|t|}
        -A_-\frac{1-\mathrm{sgn}(t)\gamma_4}{2}e^{-M_-|t|} 
  \label{eqn:asy_form_of_cor}
\end{equation}
at large Euclidean time $t$. 
Here $A_\pm$ and $M_\pm$ denote the amplitude and mass 
of the lowest-lying state in each parity channel, respectively. 

On the lattice with finite extent $T$ in the time direction, 
the two-point function receives the reflection from 
the time direction boundary.
 Under the periodic/anti-periodic boundary condition, 
Eq. (\ref{eqn:asy_form_of_cor}) should be expressed as 
\begin{eqnarray}
  G_B^\mathrm{p.b.c/a.p.b.c.}(t) &\simeq &
         \left[A_+\frac{1+\gamma_4}{2}e^{-M_+t}
              -A_-\frac{1-\gamma_4}{2}e^{-M_-t}\right]  \nonumber \\
  && \pm \left[A_+\frac{1-\gamma_4}{2}e^{-M_+(T-t)}
              -A_-\frac{1+\gamma_4}{2}e^{-M_-(T-t)}\right], 
\end{eqnarray}
where only the primal reflections are included since 
the higher orders of wrap-round effect are negligible in comparison
with the first wrap-round effect. 
Clearly, unwanted-parity contaminations originated 
from the first wrap-round effect cannot be removed by 
the simple projection 
$\mathrm{Tr}\left[\frac{1\pm \gamma_4}{2}
G_B^\mathrm{p.b.c./a.p.b.c.}(t)\right]$.
To extract the state with desired parity, 
we employ the linear combination of two correlation functions 
with periodic and anti-periodic boundary conditions, 
$\bar{G}_B(t)=
\{G_B^\mathrm{p.b.c.}(t)+G_B^\mathrm{a.p.b.c.}(t)\}/2$.
As the results, the free-boundary situation is realized, 
so the simple projection becomes efficient. 
In the numerical simulation, 
we adopt a linear combination of the quark propagators. 
As shown in Ref. \cite{Sasaki:2005ug}, 
the linear combination in the quark level automatically realizes 
the linear combination in the hadronic level. 
We perform this parity projection to apply to baryon two-point correlators. 
We access to the two different parity states of the nucleon, 
which correspond to 
the positive-parity state ($N$) and 
the negative-parity state ($N^*$).


We also treat the $\Delta$ baryon in this study. 
The simplest composite operator for the $\Delta$ baryon
${\cal O}_i^\Delta(x)=\varepsilon_{abc}
(u_a^T(x)C\gamma_i u_b(x))u_c(x)$ 
has the structure of the Rarita-Schwinger spinor 
so that the two-point function constructed from this operator 
coupled to both the spin-3/2 and spin-1/2 state \cite{Leinweber:1992hy}. 
Here $i$ is the spatial Lorentz index. 
The two-point function 
$G_{ij}^\Delta(t)=\sum_{\vec{x}}\langle 0|
T\{{\cal O}_i^\Delta(\vec{x},t)\bar{\cal O}_j^\Delta(\vec{0},0)\}
|0\rangle$ is expressed by the orthogonal sum of 
spin-3/2 and spin-1/2 components: 
\begin{equation}
  G_{ij}^\Delta(t)=
  \left(\delta_{ij}-\frac{1}{3}\gamma_i\gamma_j\right)G_{3/2}^\Delta(t) 
  +\frac{1}{3}\gamma_i\gamma_j G_{1/2}^\Delta(t) 
\end{equation}
with appropriate spin projection operators \cite{Benmerrouche:1989uc}. 
Therefore, we can access the excited $\Delta$ state 
such as the spin-1/2 $\Delta$ state by the spin projected correlators 
\begin{eqnarray}
  G_{3/2}^\Delta(t) &=& \frac{3}{2}G_{ii}^\Delta(t)
  -\frac{1}{2}\sum_k \gamma_i\gamma_k G_{ki}^\Delta(t), \\
  G_{1/2}^\Delta(t) &=& \sum_k \gamma_i\gamma_k G_{ki}^\Delta(t).
\end{eqnarray}
We perform the appropriate parity projection 
to apply to $G_{3/2}^\Delta(t)$ and $G_{1/2}^\Delta(t)$ as well. 
We finally access to 
the four different spin-parity states of the $\Delta$ baryon, 
which correspond to 
the spin-3/2 positive-parity state ($\Delta_{3/2}$), 
the spin-3/2 negative-parity state ($\Delta_{3/2}^*$), 
the spin-1/2 positive-parity state ($\Delta_{1/2}$) and 
the spin-1/2 negative-parity state ($\Delta_{1/2}^*$). 
The $\Delta_{3/2}$ is the ground state of the $\Delta$.



The gauge configurations have been generated in quenched approximation 
with the single plaquette gauge action at $\beta=6/g^2=6.2$, 
where the cut-off scale ($\sim 3$ GeV) was definitely higher than 
mass scale of all calculated baryons ($\sim 1-2$ GeV). 
To examine the finite size effect, 
we performed simulations on three different lattice sizes, 
$L^3\times T=24^3\times 48~ (N_\mathrm{conf}=320)$, 
$32^3\times 48~ (N_\mathrm{conf}=240)$ and 
$48^3\times 48~ (N_\mathrm{conf}=210)$. 
To compute the quark propagator, 
we utilized the standard Wilson fermion action, 
of which computational costs are relatively cheep. 
The hopping parameter $\kappa=0.1520,~ 0.1506,~ 0.1497,~ 0.1489,~ 0.1480$ 
cover the range $M_\pi/M_\rho\simeq 0.66-0.96$. 
To perform the appropriate parity projection, 
we adopt a procedure to take an average of two quark propagators,
which are subjected to periodic and anti-periodic boundary condition in time. 
We calculated  the point-to-point quark propagator and constructed
the baryon two-point functions by the conventional interpolating operators, 
$\varepsilon_{abc}(u_a^TC\gamma_5 d_b)u_c$ for the nucleon, 
and $\epsilon_{abc}(u_a^TC\gamma_i u_b)u_c$ for the $\Delta$ baryon. 

%
%
\begin{figure}[htbp]
  \begin{center}
  \includegraphics[height=60mm]{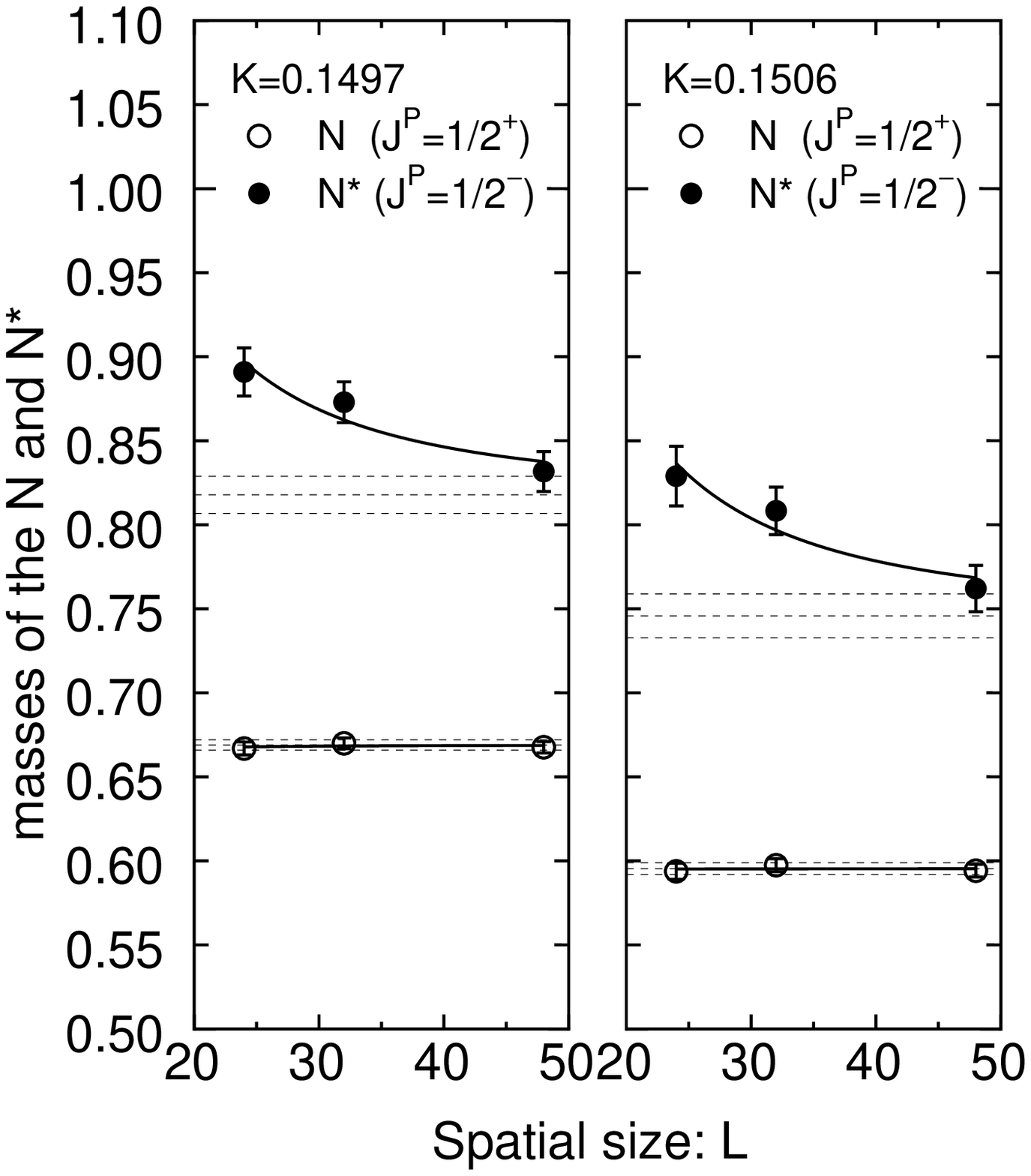}
  \hspace{25mm}
  \includegraphics[height=60mm]{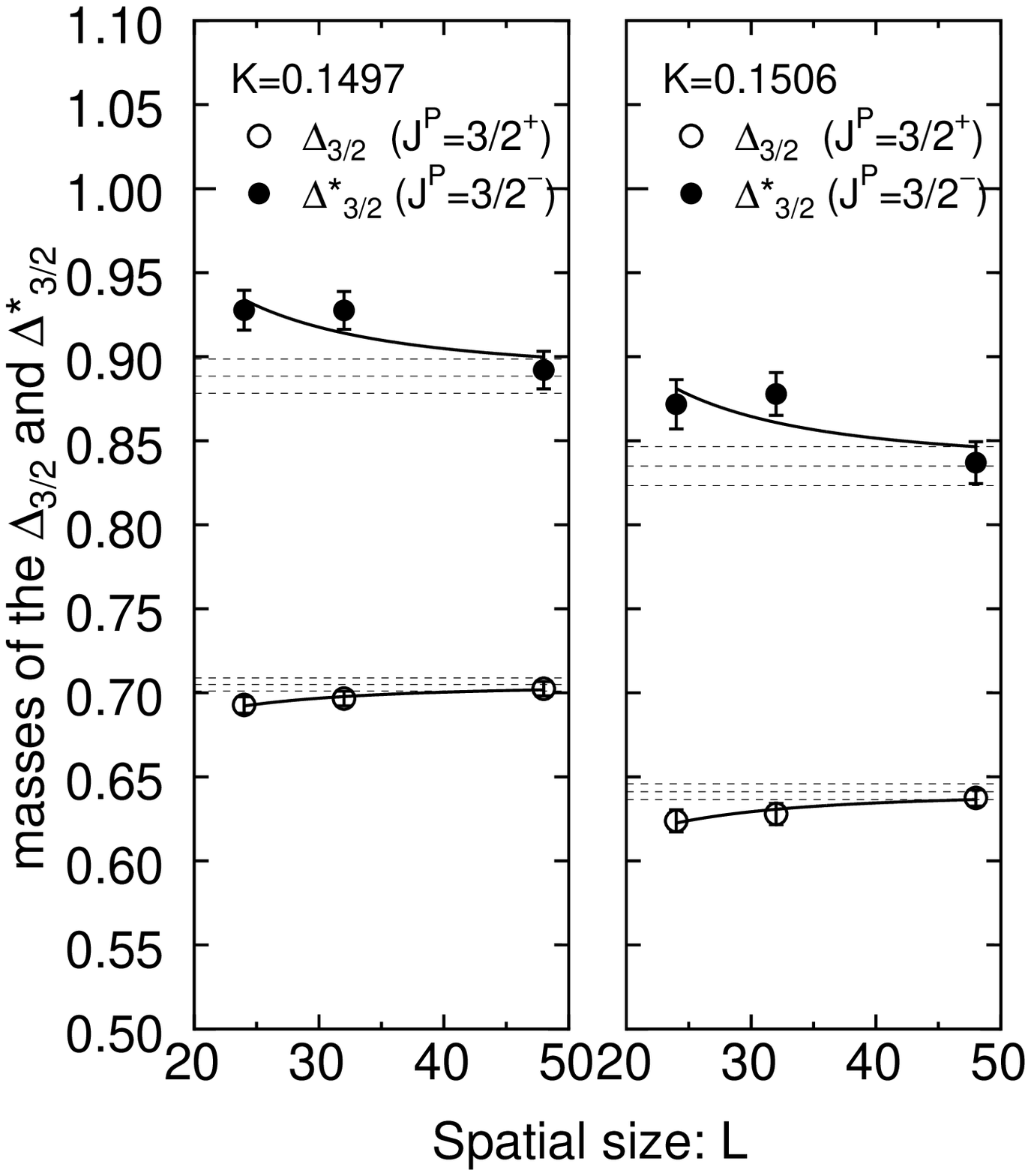}
  \end{center}
  \caption{Masses of the $J^P=1/2^\pm$ nucleons (left figures) 
  and the $J^P=3/2^\pm$ $\Delta$ baryons (right figures) 
  as function of spatial lattice size in lattice units 
  for two hopping parameters ($\kappa=0.1497$ and $0.1506$).}
  \label{fig:size_effect_of_baryon_mass}
\end{figure}

In Fig.\ref{fig:size_effect_of_baryon_mass}, 
the lattice-size dependence on the mass of each baryon 
($N$, $N^*$, $\Delta_{3/2}$, $\Delta_{3/2}^*$) 
is shown for two hopping parameters ($\kappa=0.1506$ and $0.1497$), 
which correspond to the relatively heavier quark masses. 
The quoted errors represent only statistical errors. 
Horizontal dashed lines represent the values in the infinite volume limit 
together with their one standard deviation. 
Those values are evaluated with the phenomenological power-low formula, 
$aM_L=aM_\infty +cL^{-2}$.

For the ground-state of the nucleon, we do not
observe any serious finite size effect
even on our smallest lattice. 
On the other hands, 
we find that the mass of the negative-parity nucleon 
suffers from the large finite size effect. 
The observed tendency of the size effect is that
the mass decreases as the lattice size increases.
This result seems to be consistent with the picture 
where the squeezed ``wave function'' may increase the total energy
of the three-quark system and the ``wave function'' of the excited state 
has the larger spatial extent than that of the ground state.

For the $\Delta_{3/2}$ baryon, however, 
we find that the finite size effect has a different pattern:
$M_{\Delta_{3/2}}$ becomes large as the lattice size increases.
On the other hands, 
the $\Delta_{3/2}^*$ has the same pattern of the finite size effect 
observed in the $N^*$ spectrum, 
while the finite size correction of the $\Delta_{3/2}^*$ 
is relatively milder than that of the $N^*$. 
We confirm the similar feature 
in the case of the $\Delta_{1/2}$ baryon as well. 
In all cases, to keep the size effect at a few percent levels, 
the spatial size of lattice is required to be at least 3 fm. 

%
%
\begin{figure}
  \begin{center}
  \includegraphics[height=60mm]{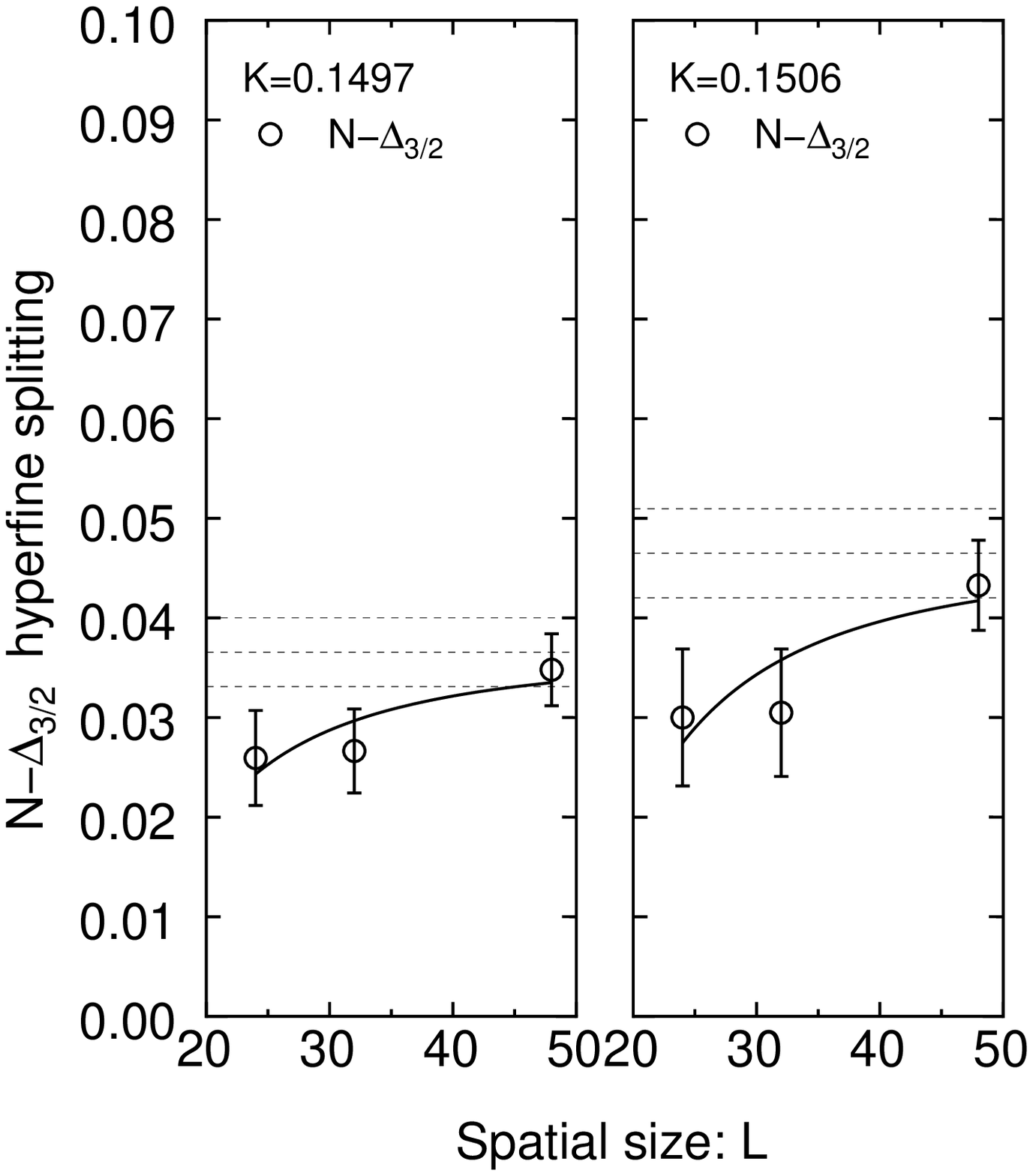}
  \hspace{25mm}
  \includegraphics[height=60mm]{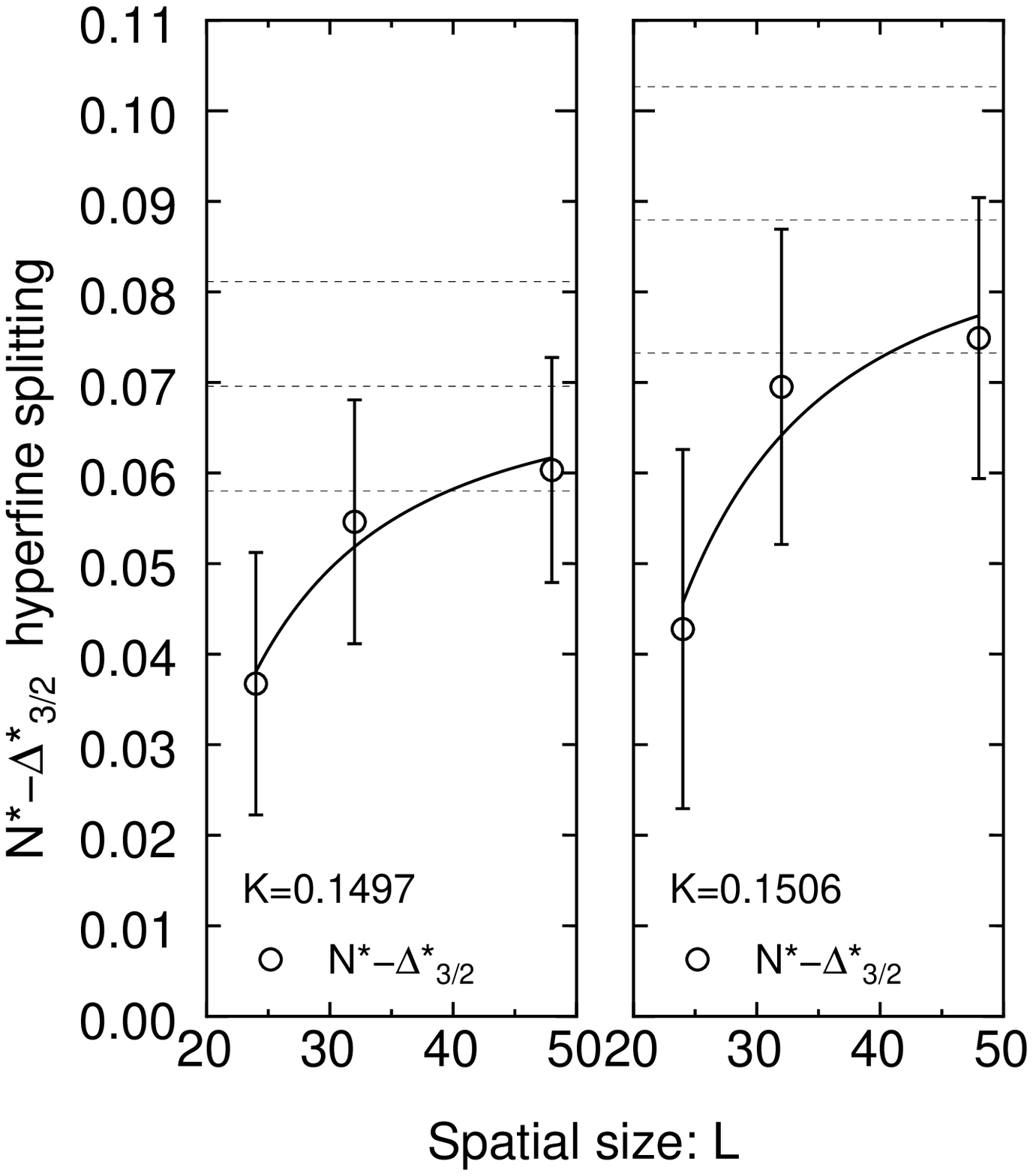}
  \end{center}
  \caption{$N-\Delta_{3/2}$ hyperfine splittings (left figures) 
  and $N^*-\Delta_{3/2}^*$ hyperfine splittings (right figures) 
  as function of spatial lattice size in lattice units 
  for two hopping parameters ($\kappa=0.1497$ and $0.1506$).}
  \label{fig:size_effect_of_hyperfine}
\end{figure}

The peculiar behavior on the $\Delta_{3/2}$ 
may originate from a hyperfine interaction. 
Our simulations are carried out in the relatively heavy quark region, 
where the size effect of the nucleon is almost negligible. 
Thus, the finite size effect on the $\Delta_{3/2}$ 
can attribute to the effect of 
the $N-\Delta_{3/2}$ hyperfine mass splitting. 
In Fig. \ref{fig:size_effect_of_hyperfine}, 
the lattice-size dependence of the hyperfine mass splitting 
in each parity channel is shown. 
In the negative-parity $N^*-\Delta_{3/2}^*$ hyperfine splitting, 
we observe the same pattern as the $N-\Delta_{3/2}$ hyperfine splitting.
The same feature is also observed 
in the charmonium spectrum~\cite{Choe:2003wx}.
The hyperfine splitting between $\eta_c$ and $J/\psi$ diminishes
as the spatial lattice size decreases if $La \le 1.3$ fm.
It seems that it is difficult for one-gluon exchange as a spin-spin 
component of the Fermi-Breit type interaction to account for
this particular pattern of the finite size effect 
on the hyperfine interaction \cite{Sasaki:2005ug}. 


Next, we extrapolate the baryon masses calculated on our largest lattice 
($La\simeq 3.2$ fm) toward the chiral limit. 
Our quark masses are too heavy 
to employ guidance by the chiral perturbation theory, 
so the empirical curve fit formula is alternatively utilized: 
\begin{equation}
  (aM_B)^2 = d_0 + d_2 (aM_\pi)^2. 
\end{equation}
As reported in some Refs. \cite{Gockeler:2001db,Sasaki:2005ug}, 
this formula gives a better fit to the data in the heavy quark region 
than the linear fit. 

In Table \ref{tab:comparison_exp}, 
the baryon masses in the chiral limit are listed in the physical unit. 
Two types of input ($r_0$ input and $M_\rho$ input) are taken 
to reveal the dependence of the choice of input to set a scale. 
Since the systematic errors stemming from this choice exceed 
the statistical errors of the nucleon or the $\Delta_{3/2}$, 
we alternatively quote various mass ratios also, 
\begin{eqnarray}
  M_{\Delta_{3/2}}/M_{N} 
  &=& 1.28(4)\hspace{5mm}(\mbox{Expt. : }\sim 1.31),
  \nonumber \\
  M_{N^*}/M_{N} 
  &=& 1.61(10)\hspace{3.1mm}(\mbox{Expt. : }\sim 1.63),
  \nonumber \\
  M_{\Delta_{3/2}^*}/M_{N^*} 
  &=& 1.28(7)\hspace{5mm}(\mbox{Expt. : }\sim 1.11),
  \nonumber \\
  M_{\Delta_{3/2}^*}/M_{\Delta_{3/2}} 
  &=& 1.61(7)\hspace{5mm}(\mbox{Expt. : }\sim 1.38).
\end{eqnarray}
The mass ratios show a good agreement with 
the experimental values within statistical errors 
except for the case where $M_{\Delta_{3/2}^*}$ is included.
The values extrapolated from large quark mass should not be taken seriously. 
However, the level ordering in $\Delta$ spectra, 
$M_{\Delta_{3/2}}<M_{\Delta_{1/2}^*}\simle
M_{\Delta_{3/2}^*}<M_{\Delta_{1/2}}$, 
is well-reproduced in comparison to corresponding $\Delta$ states, 
which are all ranked as four stars on the Particle Data Table 
\cite{Eidelman:2004wy}. 
In addition, it is worth mentioning that 
a signal $\Delta(1750)$ $(I=3/2\mbox{ and }J^P=1/2^+)$, 
which is the weakly established state (one star) 
\cite{Eidelman:2004wy}, cannot be seen in our data.

\begin{table}[htbp]
\caption{The third and forth column list results of all measured
baryon masses in GeV units, which are set by two different inputs, 
$r_0$ input and $M_\rho$ input. The fifth and sixth list experimental 
values of the corresponding baryon and its status 
in the Particle Data Table \cite{Eidelman:2004wy}.
The possible assignments of the $SU(6)\otimes O(3)$ supermultiplet 
are also embedded into the final column. }
\begin{center}
\begin{tabular}{cc|ll|l|c}
\hline
\hline
baryon &($I,J^P$) & our results  [GeV]  &  & physical state (status) & $SU(6)\otimes O(3)$ \\
&& ($r_0$ input)& ($M_{\rho}$ input) & &classification \\
\hline
$N$ & 
$(\frac{1}{2},\frac{1}{2}^{+})$ & 0.968(23) &  1.031(25) &$N(939)$ **** & [{\bf 56},$0^+$]\\
$N^{*}$&
$(\frac{1}{2},\frac{1}{2}^{-})$  & 1.555(97) &  1.658(102)&$N(1535)$ $S_{11}$  **** & [{\bf 70},$1^-$]\\
$\Delta_{3/2}$&
$(\frac{3}{2},\frac{3}{2}^{+})$ & 1.236(29) &  1.320(31)&$\Delta(1232)$ $P_{33}$ **** & [{\bf 56},$0^+$]\\
$\Delta^{*}_{3/2}$&
 $(\frac{3}{2},\frac{3}{2}^{-})$ & 1.985(72) &  2.114(75)&$\Delta(1700)$ $D_{33}$ **** & [{\bf 70},$1^-$]\\
$\Delta_{1/2}$&
$(\frac{3}{2},\frac{1}{2}^{+})$ & 2.325(112) & 2.478(118)&$\Delta(1910)$ $P_{31}$ **** & [{\bf 56},$2^+$] or [{\bf 70},$0^{+}$]\\
$\Delta^{*}_{1/2}$&
 $(\frac{3}{2},\frac{1}{2}^{-})$ & 1.866(153) & 1.987(165)&$\Delta(1620)$ $S_{31}$ ****&[{\bf 70},$1^-$]\\
\hline
\hline
\end{tabular}
\label{tab:comparison_exp}
\end{center}
\end{table}


In this study, we have investigated
the finite size effect on masses of the ground and excited baryons, 
the $J^P=1/2^\pm$ nucleons and the $J^P=1/2^\pm, 3/2^\pm$ $\Delta$ baryons. 
For this purpose, we perform numerical simulations at
three different lattice sizes, $La\simeq 1.6$, 2.2 and 3.2 fm. 
We have found the sizable finite size effect on masses of 
all $\Delta$ states and the negative-parity nucleon 
even in the heavy quark region ($M_\pi/M_\rho\simeq 0.82,~ 0.87$),
where that on the nucleon is negligible. 
To keep this effect at a few percent levels, 
the spatial size of lattice is required to be at least 3 fm. 
We found a peculiar lattice-size dependence  
on the mass of the positive-parity $\Delta_{3/2}$ state:
the mass of the $\Delta_{3/2}$ state becomes small 
as the lattice size decreases.
This behavior seems to originate from 
the finite size effect on the hyperfine interaction.
We also extrapolate the baryon masses toward the chiral limit 
with the empirical curve-fit formula. All mass spectra of the 
$J^P=1/2^\pm$ nucleons and the $J^P=1/2^\pm, 
3/2^\pm$ $\Delta$ baryons are roughly consistent with 
experimental values.


\end{document}